%% file: Gravitational_polarization_1D_Polytropes.tex
\documentclass[final,5p,times,twocolumn,authoryear]{elsarticle}

\usepackage{amsmath}
\usepackage{amsfonts}
\usepackage{amssymb}
\usepackage[titletoc,title]{appendix}
\usepackage{booktabs}
\usepackage[mmddyyyy]{datetime}
\usepackage{dutchcal}
\usepackage{enumitem}
\usepackage{graphicx}
\usepackage{hyperref}
\usepackage[utf8]{inputenc}
\usepackage{lineno}
\usepackage[percent]{overpic}
\usepackage{pgfplots}
\usepackage{relsize}
\usepackage{subfigure}
\usepackage{siunitx}
\usepackage{tikz}

\modulolinenumbers[5]
\numberwithin{equation}{section}

\usetikzlibrary{calc}
\usetikzlibrary{plotmarks}
\usetikzlibrary{positioning}
\usetikzlibrary{fit}
\usetikzlibrary{decorations.pathmorphing}
\usetikzlibrary{backgrounds}
\pgfplotsset{compat = newest}
\pgfplotsset{ legend style={font=\tiny} }
\definecolor{bgreen}{rgb}{0.0,0.5,0.0}
\definecolor{bblue}{rgb}{0.0,0.0,0.9}
\definecolor{bgold}{rgb}{0.7,0.5,0.0}
\definecolor{bred}{rgb}{0.9,0.0,0.0}

\begin{document}

\begin{frontmatter}



\title{Gravitational polarization of test-mass potential in equilibrium polytropic sheets with non-negative polytropic indexes}


\author{Yuta Ito}
\ead{yutaito30@gmail.com}

\begin{abstract}
Gravitational polarization is examined for equilibrium self-gravitating polytropic sheets perturbed by gravitational field due to test mass sheet. We find equilibrium solutions to the corresponding perturbed Lane-Emden equations for non-negative polytropic indexes. It is shown that gravitational polarization may be observed even in a \emph{finite} extent of self-gravitating systems in addition to previously discussed infinite systems. In the polytropic sheets, the maximum gravitational amplification gets greater with a higher polytropic index while the height at which the maximum amplification occurs gets lower. The ratio of height change to the original height increases with polytropic index. The last result constrains the linear approximation method used for the present perturbation method.  

\end{abstract}


\end{frontmatter}

\input{Section1_intro.tex}

\input{Section2_method_poly_sheet.tex}

\input{Section3_n_01_poly_sheet.tex}

\input{Section4_numerical_results.tex}
\input{Section5_conclusion.tex}



\bibliographystyle{elsarticle-harv} 
\bibliography{science}


\end{document}

%% file: Section1_intro.tex
\section{Introduction}

Gravitational polarization of fields induced by test mass is an important but little-explored fundamental concept in statistical dynamics of self-gravitating systems. Polarization effects are generally known for dielectrics under external electric fields and for electrolytes or plasmas perturbed by a test-charge potential. In the systems, positive and negative charge distributions are reconfigured so that applied electric fields are weaken or even shielded. On the one hand, if test mass is added in a self-gravitating system, gravitational field due to test mass could be reinforced \citep{Miller_1966,Gilbert_1970}. The mechanism is simple; gravitational field due to test mass attracts ambient masses toward the test mass, the reconfigured mass distribution then can strengthen the field. Gravitational polarization could help understand the fundamental response of self-gravitating system to external perturbations. Examples are a deformation of polarized halo due to disc growth \cite{Murali_1998, Moody_1999}, perturbations in massive haloes \citep{Murali_1999},  stellar accretion and heavy central astrophysical objects \citep{Young_1980, Murali_1998, Quinlan_1995}, interaction between ``dressed'' particles/stars \citep{Heyvaerts_2010, Chavanis_2012}, and flyby of heavy objects \citep{Vesperini_2000}. These settings are realistic but complicated, so they are not suitable to assess \emph{pure} effects and properties of gravitational polarization. The simplest setting is to examine gravitation amplification is to add or move test mass to the center of gravity in a self-gravitating system so that the spatial symmetry is held through test-mass perturbation \citep{Gilbert_1970, Goodman_1984a}.

Discussing gravitational polarization of point-mass potential, however, is not a straightforward topic. First, self-gravitating systems are inhomogeneous and finite in size because of self-gravity. We can not resort to an infinite homogeneous approximation, which often allows us to use simple mathematical deductions. Such an approximation was applied to collisionless stellar systems \citep{Marochnik_1968} and cold dark matters \citep{Padmanabhan_1985} with the Maxwellian distribution function for stars and particles. The works suggested that gravitational potential due to test point mass behaves like $\cos(k_\text{J}r)/r$, where $k_\text{J}$ is the Jeans wave number, and $r$ the distance from test mass. The configuration, however, suffers from the problem of ``Jeans swindle'' \citep[e.g.,][]{Falco_2013} which is an inconsistent method unless we consider a cosmological setting. Second, particles and stars orbit obeying self-consistent Newtonian mean-field potentials in self-gravitating systems. This feature is especially important for stars in stellar systems, such as galaxies, globular clusters, and nuclear star clusters. Stars move around forming \emph{smooth} orbits that are not disturbed irregularly by other stars on dynamical time scales \citep[e.g.,][]{Binney_2011}. This implies that we need to know explicitly analytical expressions to describe orbits, such as orbital periods and isolating integrals. It is possible to obtain the expressions only for limited cases. \cite{Goodman_1984a} used a harmonic oscillator as an approximation of stellar cluster core. A successful discussion of gravitational amplification was made for a collisionless isochrone \citep{Gilbert_1968} where an emerging test-point mass is placed at the center of the isochrone holding the conservation of total mass. Gilbert showed that gravitational field due to test mass is strengthened with radius at small radii and reaches its maximum once. At larger radii, the effect of amplification disappears as the density approaches zero. Third, even with extensive numerical methods, the orbital effects can be handled only for weak perturbation. To overcome the problem of orbital effects above,  it is possible to use \cite{Kalnajs_1977}'s matrix method for realistic self-gravitating systems, such as the King models. However, the method applies only to weak perturbation problems. Even for such limited setting, the method needs exhaustive series expansions with sophisticated numerical schemes \citep{Murali_1999}. It appears that no further works are found with this method for gravitational polarization.  

For this situation, we recently discussed gravitational polarization in infinite collisional gaseous systems, namely the isothermal sheets, cylinder, and sphere \citep{Ito_2023a}. There are advantages in examining the models. (i) We can avoid the orbital effects of particles because of high collisionality in the systems. (ii) As discussed in \citep{Murali_1998}, gaseous (fluid) system could show similar response features to collisionless systems. (iii) Not only linear but non-linear analyses are easily executed because of simple mathematical structures of the Lane-Emden equation. In \citep{Ito_2023a}, we found in the isothermal sheets and cylinder that potentials due to test masses show the same qualitative characteristics as that in collisionless isochrone. On the one hand, the isothermal sphere showed an oscillatory gravitational field due to test point mass similar to that in infinite homogeneous collisionless systems with the Maxwellian distribution. The isothermal systems seems \emph{nice} models to discuss gravitational polarization compared to the previous results.  Hence, it would be reasonable to explore other collisional self-gravitating systems as well for further discussions.

All the previous works focusing on gravitational polarization examined only \emph{infinite} self-gravitating systems. Such systems are unrealistic in nature. It is important to examine whether gravitational polarization may be observed even in \emph{finite} systems as well. The present paper examines gravitational polarization in equilibrium self-gravitating polytropic sheets. The mass density and height of the sheet models are finite in the stratified direction for polytropic index of $0\leq n < \infty$ \citep[e.g.,][]{Horedt_2000}. The models are important to understand stratification and fragmentation of self-gravitating gaseous systems.  Also, they are stable against radial perturbation. The last feature may be less attractive for a statistical-dynamics point of view \citep{Campa_2009} since the sheet models do not show exotic collective features, such as negative specific heat. Yet, we believe that excluding those features can make easier our understanding of gravitational polarization. 

The goal of the present work is to show gravitational polarization in polytropic sheets. We assume that test mass sheet is placed perpendicularly to the stratified direction at the center of the sheets. We numerically solve a linearlized Lane-Emden equation perturbed by potential due to test mass. To account the finiteness of system size, we employ the method used for a tidal effect on polytropic spheres \citep{Chandrasekhar_1933a,Chandrasekhar_1933b}. Our numerical results show that the maximum of gravitational field gets greater with a higher polytropic index in polytropic sheets. In the limit $n>>1$, the maximum approaches that of the isothermal polytrope. We also show that the shortening of polytropic sheets is more significant for higher polytropic indexes. This result provides the limit of the linearlization approximation that we use for test-mass perturbation.

The present paper is organized as follows. Section \ref{sec_poly_models} explains polytropic sheet models and the perturbation method to examine gravitational polarization. Section \ref{Sec_analy} shows analytical  results for polytropes of $n=0$ and $n=1$ to which our perturbation method is applied. It also introduces measures of gravitational amplification and shortening of polytropic sheets. Section \ref{sec:result} shows numerical results for $0\leq n<\infty$. Section \ref{sec:conclusion} is Conclusion.

%% file: Section2_method_poly_sheet.tex
\section{Polytropic sheets and its perturbation due to test mass}\label{sec_poly_models}

The present section first explains equilibrium polytropic sheets and a perturbation method for the sheets due to test mass sheet. It then explains the effect of shortening on the sheets.

\subsection{Deriving the Lane-Emden equation for polytropic sheets}\label{Deriv_Poly_1D}

Polytropic sheet models are three-dimensional self-gravitating equilibrium model composed of particles interacting via the pair-wise Newtonian potential \citep[e.g.,][]{Horedt_2000}. The self-consistent mean-field potential $\Phi(\boldsymbol{r})$ is determined by mass density $\rho(\boldsymbol{r})$ via the Poisson equation
\begin{equation}
	\frac{\partial}{\partial \boldsymbol{r}}\cdot\left(\frac{\partial \Phi}{\partial \boldsymbol{r}}\right)=4\pi G\rho(\boldsymbol{r}),\label{Eq_Pois}
\end{equation}
where $G$ is the gravitational constant. In the equilibrium model, the hydrostatic equation must hold;
\begin{equation}
	\frac{\partial p}{\partial \boldsymbol{r}}+ \rho(\boldsymbol{r})\frac{\partial \Phi}{\partial \boldsymbol{r}}=0,\label{Eq_hydro}
\end{equation}
where $p(\boldsymbol{r})$ is the isotropic scalar pressure. We assume polytropic sheets to satisfy the polytropic relation
\begin{equation}
	p=K\rho^{1+1/n}, \label{Eq_poly}
\end{equation}
where $K$ is the polytropic constant and $n$ the polytropic index. Equations \eqref{Eq_Pois},  \eqref{Eq_hydro}, and \eqref{Eq_poly} are a set of closed equations for variables $\rho(\boldsymbol{r})$, $p(\boldsymbol{r})$, and $\Phi(\boldsymbol{r})$. 

With dimensionless variables
\begin{eqnarray}
	      \phi&\equiv&\frac{\Phi(\boldsymbol{r})}{K(n+1)\,\rho_\text{c}^{1/n}},\\
	\boldsymbol{\xi}&\equiv&\frac{\boldsymbol{r}}{L_\text{c}}\equiv\left(\frac{4\pi G}{K(n+1)}\rho_\text{c}^{1-1/n}\right)^{1/2}\boldsymbol{r},\label{dimless_var}
\end{eqnarray}
where $\rho_\text{c}$ is the central density of $\rho(r)$, equations \eqref{Eq_Pois},  \eqref{Eq_hydro}, and \eqref{Eq_poly} reduce to the Lane-Emden equation for polytropes of index $n$
\begin{equation}
	\frac{\partial}{\partial \boldsymbol{\xi}}\cdot\left(\frac{\partial \phi}{\partial \boldsymbol{\xi}}\right)-\left(-\phi\right)^n=0,\label{g_LE}
\end{equation}
where the following relations are used
\begin{eqnarray}
	\rho&=&\rho_\text{c}\left(-\phi\right)^n,\\
	   p&=&K\rho_\text{c}^{1+1/n}\left(-\phi\right)^{n+1}.
\end{eqnarray}

\begin{figure}
	\centering
	\begin{tikzpicture}[scale=0.9]
		\begin{semilogxaxis}[ grid=major, xmax=5, xmin=3e-4, xlabel=\Large{$z$},ylabel=\Large{$\rho$}, legend pos=south west]
			\addplot [color = red ,mark=no,thick,solid ] table[x index=0, y index=1]{Poly1D_n1_r_rho_DiffRho.txt}; 
			\addlegendentry{\large{$n=1$}}
			\addplot [color = orange ,mark=no,thick,densely dashed ] table[x index=0, y index=1]{Poly1D_n2_r_rho_DiffRho.txt}; 
			\addlegendentry{\large{$n=2$}}
			\addplot [color = purple ,mark=no,thick,densely dotted ] table[x index=0, y index=1]{Poly1D_n5_r_rho_DiffRho.txt}; 
			\addlegendentry{\large{$n=5$}}
			\addplot [color = blue ,mark=no,thick,dashed ] table[x index=0, y index=1]{Poly1D_n20_r_rho_DiffRho.txt}; 
			\addlegendentry{\large{$n=20$}}
			\addplot [color = black ,mark=no,thick,dotted ] table[x index=0, y index=1]{Poly1D_n100_r_rho_DiffRho.txt};
			\addlegendentry{\large{$n=100$}}
		\end{semilogxaxis}
	\end{tikzpicture}
	\caption{Mass density distribution of the polytropic sheet models.}
	\label{fig:1D_rho}
\end{figure}
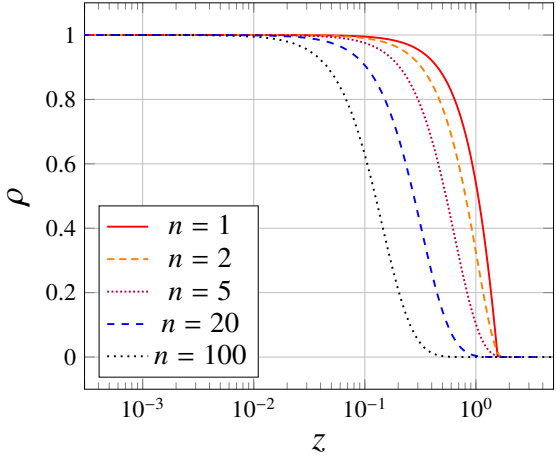

The polytropic sheets are stratified in a direction, say the $z$-direction. Equation \eqref{g_LE} for stratified sheets along the $z$-axis is written as
\begin{equation}
	\frac{\text{d}^2\phi}{\text{d}\,z^2}-\left(-\phi\right)^n=0.
	\label{LE_1D}
\end{equation} 
The boundary conditions (BCs) for equation \eqref{LE_1D} are 
\begin{equation}
	\phi(z=0)=-1, \qquad \phi'(z=0)= 0. \label{BC_LE_1D}
\end{equation}
Our interest is polytrope sheets of $0\leq n<\infty$ whose potential monotonically decreases with height and reaches zero at a finite height. We label the maximum height, or the first zero of the potential, as $z_\text{M}$ hereafter. Their mass density also has the same characteristics as the potential except for $n=0$; the density is uniform in the polytrope of $n=0$. Figure \ref{fig:1D_rho} shows mass density distributions at $z>0$ for selected polytropic indexes.
 
\subsection{Perturbation method for the polytropic sheets}\label{Perturb_Poly1D}

The present mathematical formulation for gravitational amplification is the same as that for gravitational tides \citep{Chandrasekhar_1933b} except for the location of test mass. The \emph{bare} gravitational field strength due to test mass sheet is constant. Hence, even if it is placed outside a polytrope, it does not cause tidal effect. On the one hand, if it is placed at the center of gravity of the polytrope, it causes gravitational amplification. 

We use the perturbation method developed for the isothermal systems \citep{Ito_2023a}. Imagine that test mass sheet is added at $z=0$ on the $xy$-plane in a polytropic sheet model. Equation \eqref{LE_1D} is modified as
\begin{equation}
	\frac{\text{d}^2\phi}{\text{d}\,z^2}-\left(-\phi\right)^n=\epsilon\delta(z),
	\label{LE_1D_perturb}
\end{equation} 
where $\epsilon$ is a small parameter defined as
\begin{equation}
	\epsilon\equiv\dfrac{m_\text{p}}{\rho_\text{c}L_\text{c}},
\end{equation}
where $m_\text{p}$ is test mass.  Expand equation \eqref{LE_1D_perturb} with $\epsilon$ as
\begin{equation}
	\frac{\text{d}^2}{\text{d}\,z^2}\left(\phi_\text{o}+\epsilon\delta\phi\right)-\left(-\phi_\text{o}-\epsilon\delta\phi\right)^n=\epsilon\delta(z),
	\label{LE_1D_perturb_expan}
\end{equation} 
where $\phi_\text{o}$ is the unperturbed potential of polytropic sheet model and $\epsilon\delta\phi$ the potential deviated from $\phi_\text{o}$ due to test mass. At the order of $\epsilon$, we have a linearized Lane-Emden equation for $\delta\phi$;
\begin{equation}
	\frac{\text{d}^2 \delta\phi}{\text{d}\,z^2}+n\left(-\phi_\text{o}\right)^{n-1}\delta\phi=\delta(z).
	\label{LE_1D_perturb_expan_delPhi}
\end{equation} 

We introduce the following potential
\begin{equation}
	W(z)\equiv\delta\phi(z)-\frac{1}{2}|z|,\label{Eq.W}
\end{equation}
so that equation \eqref{LE_1D_perturb} does not include the delta function 
\begin{equation}
	\frac{\text{d}^2 W}{\text{d}\,z^2}+n\left(-\phi_\text{o}\right)^{n-1}\left(W+\frac{1}{2}|z|\right)=0.
	\label{LE_1D_perturb_reg}
\end{equation}
With equation \eqref{Eq.W}, there is no gravitational field due to $W(z)$ at $z=0$. Hence, the BC is
\begin{equation}
	W'(z=0)=0.\label{BC_1D_x_0}
\end{equation}
Another BC is determined so that the effective mass $m^{*}$, the partial sum of reconfigured sheet masses, is zero at $|z|=z_\text{M}$. The effective mass for the polytropic sheet is at $z$
\begin{eqnarray}
	m^{*}(z)&=&-2\int_{0}^{z}n\left(-\phi_\text{o}\left(z'\right)\right)^{n-1}\left(W\left(z'\right)+\frac{1}{2}|z'|\right)\text{d}z',\nonumber\\
	&=&2\left(W'(z)-W'(0)\right).
\end{eqnarray}
The mass $m^{*}(z)$ must be zero at $z=z_\text{M}$ since test mass $m_\text{p}$ is not included in the reconfigured masses. Accordingly, the BC is at $|z|=z_\text{M}$
\begin{equation}
	W'(|z|=z_\text{M})=0.\label{BC_1D_x_inf}
\end{equation}
Our fundamental numerical strategy is to find potential $W(z)$ by solving equation \eqref{LE_1D_perturb_reg} with two BCs \eqref{BC_1D_x_0} and \eqref{BC_1D_x_inf} after finding the unperturbed potential $\phi_\text{o}$ by solving equation \eqref{LE_1D} with BCs in equation \eqref{BC_LE_1D}.

\subsection{Polytropic sheet shortening and surface boundary condition}

Gravitational potential due to test mass sheet pulls polytropic sheets in the $z$-direction toward the test mass, which results in a shortening of sheet thickness. The discussion of polytrope shortening is essentially the same as that of distorted polytrope due to tidal effects \citep{Chandrasekhar_1933b}. We first introduce the new maximum height of polytropic sheets due to perturbation
\begin{equation}
	z_\text{s}\equiv z_\text{M}+\epsilon\delta z,\label{Eq.zs}
\end{equation} 
where $\delta z$ is the deviation from $z_\text{M}$ at order of $\epsilon$ and expected to take a negative value. Expand the new equilibrium potential
\begin{equation}
	\phi(z)=\phi_\text{o}(z)+\epsilon\delta\phi(z). \label{Eq.phi_expan}
\end{equation}
around at $z_\text{M}$ using equation \eqref{Eq.zs} as follows  
\begin{equation}
	\phi(z_\text{s})=\phi_\text{o}(z_\text{M})+\epsilon\left(\phi_\text{o}'(z_\text{M})\,\delta z +\delta\phi(z_\text{M})\right)+\mathcal{O}\left(\epsilon^2\right).
\end{equation}
The new equilibrium potential $\phi(z)$ must be zero at $z_\text{s}$. Since $\phi_\text{o}(z_\text{M})$ is zero, we have the condition
\begin{equation}
	\delta z=-\frac{\delta\phi(z_\text{M})}{\phi_\text{o}'(z_\text{M})}=-\frac{W(z_\text{M})+\frac{1}{2}z_\text{M}}{\phi_\text{o}'(z_\text{M})}.\label{Eq.delta_z}
\end{equation}
Using equation \eqref{Eq.zs}, we next expand the derivative of potential $\phi(z)$
\begin{equation}
	\phi'(z)=\phi_\text{o}'(z)+\epsilon\delta\phi'(z)
\end{equation}
to the order of $\epsilon$ as follows
\begin{equation}
	\phi'(z_\text{s})=\phi_\text{o}'(z_\text{M})+\epsilon\left(\phi_\text{o}''(z_\text{M})\,\delta z +\delta\phi'(z_\text{M})\right).
\end{equation}
With function $W(z)$ in equation \eqref{Eq.W}, another condition is given on the surface of the perturbed polytrope as
\begin{equation}
\phi_\text{o}''(z_\text{M})\,\delta z +W'(z_\text{M})=0.\label{Eq.delta_x2}
\end{equation}
With equation \eqref{Eq.delta_z}, we hence obtain the boundary condition on the surface
\begin{equation}
	-\frac{W(z_\text{M})+\frac{1}{2}z_\text{M}}{\phi_\text{o}'(z_\text{M})}\,\phi_\text{o}''(z_\text{M}) +W'(z_\text{M})=0.\label{Eq.gener_BC}
\end{equation}
Equation \eqref{Eq.gener_BC} is a generalized BC of that in equation \eqref{BC_1D_x_inf}. For polytropes of $1<n<\infty$, the term $\phi_\text{o}''(z_\text{M})$ is zero because of equation \eqref{LE_1D}. This means that the boundary condition on a shortened polytrope surface is not affected by the shortening, or $\delta z$. The effect of shortening should appear in distorted mass distribution when $\phi_\text{o}''(z_\text{M})$ is non-zero. Such an example is that polytropic sheets are inserted between two pressurized mediums. Another example is that mass density does not reach zero at $z_\text{M}$, which corresponds with the polytropic sheet of $n=0$ in the present work.

%% file: Section3_n_01_poly_sheet.tex
\section{Gravitational polarization in the polytropic sheet models of $n=0$ and $n=1$ }\label{Sec_analy}

The present section explains gravitational amplification of polytropic sheets of $n=0$ and $n=1$. We provide the explicitly analytical form of the amplification for the models using the perturbation method explained in Section \ref{sec_poly_models}. We also introduce measures of gravitational amplification and shortening of polytropic sheets.

\subsection{Gravitational polarization for $n=0$}\label{sec:n_0} 

The Lane-Emden equation for polytrope of $n=0$ is
\begin{equation}
	\frac{\text{d}^2\phi}{\text{d}\,z^2}-1=0.
\end{equation}
With the BCs in equation \eqref{BC_LE_1D}, the solution reads
\begin{equation}
	\phi_\text{o}(z)=-1+\frac{z^{2}}{2},
\end{equation}
and the maximum height is 
\begin{equation}
	z_\text{M}=\sqrt{2}.
\end{equation}
The equation for the perturbed polytrope of $n=0$ is the Laplace equation
\begin{equation}
	W''(z)=0,
\end{equation}
with the BCs  \eqref{BC_1D_x_0} and \eqref{Eq.gener_BC}. The solution reads 
\begin{equation}
	W(z)=-\frac{1}{\sqrt{2}}.
\end{equation}
This means that gravitational field due to test mass sheet is not amplified and that only the reference value of potential constant is determined as $-1/\sqrt{2}$. Also, no shortening of the polytropic sheets occurs;
\begin{equation}
	\delta z=0.
\end{equation}
With the above analysis, we confirm the consistency of the BC \eqref{Eq.gener_BC}. If the BC \eqref{BC_1D_x_inf} is used instead, $W(z)$ is undetermined.

\subsection{Gravitational polarization for $n=1$}

The Lane-Emden equation for polytropic sheet of $n=1$ is
\begin{equation}
	\frac{\text{d}^2\phi}{\text{d}\,z^2}+\phi=0.
\end{equation}
With the BCs in equation \eqref{BC_LE_1D}, the solution reads
\begin{equation}
	\phi_\text{o}(z)=-\cos z.
\end{equation}
From equation \eqref{LE_1D_perturb_reg}, the linearized Lane-Emden equation for potential $W(z)$ is
\begin{equation}
	\frac{\text{d}^2 W}{\text{d}\,z^2}+\left(W+\frac{1}{2}|z|\right)=0.\label{eq.LE_n1}
\end{equation}
With two BCs \eqref{BC_1D_x_0} and \eqref{BC_1D_x_inf}, the solution is on $z>0$
\begin{equation}
	W(z)=\frac{1}{\sqrt{2}}\sin\left(z-\frac{\pi}{4}\right)-\frac{1}{2}z.
\end{equation}
In analogy with the standard electromagnetism, we introduce the following gravitational ``susceptibility''
\begin{equation}
	\chi(z)\equiv -2\delta\phi'(z).
\end{equation}
The quantity $\chi(z)$ is the ratio of the amplified to the \emph{bare} gravitational field due to test mass.  If $\chi(z)$ is greater than one, it means that gravitational field is amplified. It appears suitable as a measure of amplification, we hence call $\chi(z)$ the \emph{gravitational amplification} of fields due to test mass hereafter. For $n=1$, we have
\begin{equation}
	\chi(z)=\sqrt{2}\cos\left(z-\frac{\pi}{4}\right).
\end{equation}
The height $z_{\chi.\text{max}}$ at which gravitational amplification reaches the maximum is $\pi/4$. 
The maximum of gravitational amplification, $\chi_\text{max}$, is $\sqrt{2}$. We also introduce the average of $\chi(z)$
\begin{equation}
	\chi_\text{ave}\equiv\frac{1}{2z_\text{M}}\int^{z_\text{M}}_{-z_\text{M}}\chi\left(z'\right)\,\text{d}z'=\frac{\delta \phi(z_\text{M})-\delta\phi(0)}{z_\text{M}}.
\end{equation}
For $n=1$, $z_\text{M}$ is $\pi/2$. So, the value of $\chi_\text{ave}$ is
\begin{equation}
	\chi_\text{ave}=\frac{4}{\pi}.
\end{equation}

We next introduce the following quantity to examine the degree of shortening of polytropic sheets
\begin{equation}
	\eta\equiv\frac{\delta z}{z_\text{M}}=-\frac{W(z_\text{M})+\frac{1}{2}z_\text{M}}{z_\text{M}\,\phi_\text{o}'(z_\text{M})}.
\end{equation}
For $n=1$, the ratio reads
\begin{equation}
	\eta=-\frac{1}{2\sqrt{2}}.
\end{equation}

The above analytical results are important. First, they are used to confirm our numerical calculation. Second, unlike other indexes, the perturbed polytrope of $n=1$ does not explicitly depend on the unperturbed density as seen in equation \eqref{eq.LE_n1}. Polytropes with $n=1$ may give a hint of \emph{pure} gravitational-polarization effect. Comparing our result to the collisionless homogeneous systems \citep{Marochnik_1968,Padmanabhan_1985}, it seems that ``pure'' gravitational polarization causes a sinusoidal amplification.

%% file: Section4_numerical_results.tex
\section{Numerical results}\label{sec:result}
The present section shows numerical results for perturbed polytropic sheets due to test mass sheet. We first explain the local characteristics of gravitational amplification and then global ones in the sheets. We lastly assess the shrinkage of sheet thickness in the $z$-direction.

\subsection{Maximum gravitational amplification and its location}\label{sec:numel_res_chi_M}

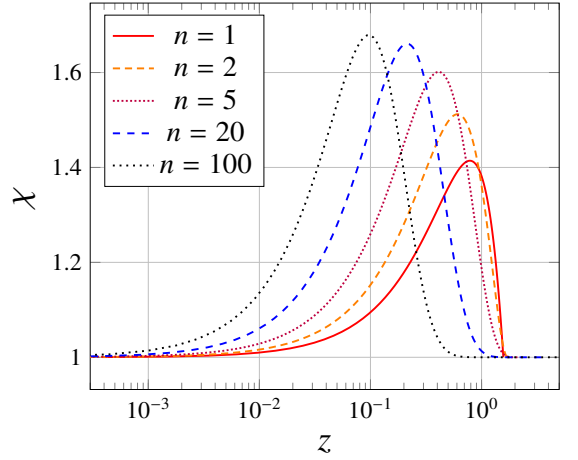
\begin{figure}
	\centering
	\begin{tikzpicture}[scale=0.9]
	\begin{semilogxaxis}[ grid=major, xmax=5e0, xmin=3e-4, xlabel=\Large{$z$},ylabel=\Large{$\chi$}, legend pos=north west]
	\addplot [color = red ,mark=no,thick,solid ] table[x index=0, y index=3]{Poly1D_n1_r_Phi_Phio_Est.txt}; 
	     \addlegendentry{\large{$n=1$}}
	\addplot [color = orange ,mark=no,thick,densely dashed ] table[x index=0, y index=3]{Poly1D_n2_r_Phi_Phio_Est.txt}; 
	      \addlegendentry{\large{$n=2$}}
	\addplot [color = purple ,mark=no,thick,densely dotted ] table[x index=0, y index=3]{Poly1D_n5_r_Phi_Phio_Est.txt}; 
	      \addlegendentry{\large{$n=5$}}
	\addplot [color = blue ,mark=no,thick,dashed ] table[x index=0, y index=3]{Poly1D_n20_r_Phi_Phio_Est.txt}; 
     	\addlegendentry{\large{$n=20$}}
	\addplot [color = black ,mark=no,thick,dotted ] table[x index=0, y index=3]{Poly1D_n100_r_Phi_Phio_Est.txt};
	     \addlegendentry{\large{$n=100$}}
	\end{semilogxaxis}
	\end{tikzpicture}
\caption{Gravitational amplification $\chi$ due to test mass sheet in the polytropic sheet models.}
\label{fig:1D_Est}
\end{figure}

Figure \ref{fig:1D_Est} depicts numerical values of gravitational amplification in selected polytropic sheets perturbed by test mass. The characteristics of $\chi(z)$ are alike among the polytropic sheets between $n=0.01$ and $n=500$. The amplification is maximized only once along the $z$-direction and weak near $z=0$ and $z=z_\text{M}$. This feature is qualitatively the same as previously reported gravitational amplification in collisionless isochrone \citep{Gilbert_1970} and isothermal cylinder and sheet \citep{Ito_2023a}. 

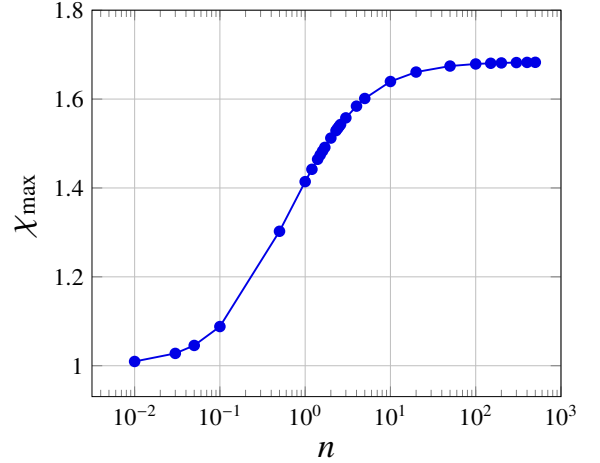
\begin{figure}
	\centering
	\begin{tikzpicture}[scale=0.9]
		\begin{semilogxaxis}[ grid=major,xlabel=\Large{$n$},ylabel=\Large{$\chi_\text{max}$},xmax=1000, ymax=1.8, legend pos=north east]
			\addplot [color = bblue ,mark=*,thick,solid ] table[x index=0, y index=3]{Poly1D_n_rM_Mtot_EstM_rEstM_EstEff_EstAVE.txt}; 
		\end{semilogxaxis}
	\end{tikzpicture}
	\caption{Maximum gravitational amplification $\chi_\text{max}$ against polytropic index $n$ for the polytropic sheet model.}
	\label{fig:1D_n_EstM}
\end{figure}

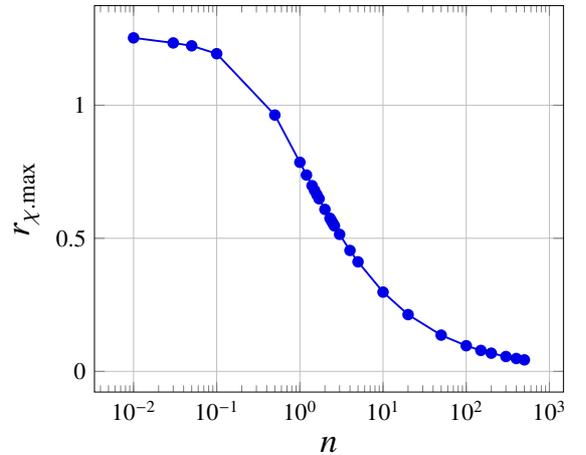
\begin{figure}
	\centering
	\begin{tikzpicture}[scale=0.9]
		\begin{semilogxaxis}[ grid=major,xlabel=\Large{$n$},ylabel=\Large{$r_{\chi.\text{max}}$}, legend pos=north east]
			\addplot [color = bblue ,mark=*,thick,solid ] table[x index=0, y index=4]{Poly1D_n_rM_Mtot_EstM_rEstM_EstEff_EstAVE.txt}; 
		\end{semilogxaxis}
	\end{tikzpicture}
	\caption{Location of $\chi_\text{max}$ against polytropic index $n$ for the polytropic sheet models.}
	\label{fig:1D_n_rEstM}
\end{figure}

In the polytropic sheets, the maximum amplification $\chi_\text{max}$ increases with index $n$ and asymptotically reaches $1.68$ (Figure \ref{fig:1D_n_EstM}). This value approximately equals the gravitational amplification obtained for the isothermal sheet model. The isothermal sheet may be considered the case of $n\to\infty$ in the polytropic sheets \citep[e.g.,][]{Horedt_2000}. On the one hand, the location of the maximum amplification gets closer to $z=0$ as $n$ increases (Figure \ref{fig:1D_n_rEstM}). This would reflect a feature of unperturbed polytropic sheets that the mass distribution is concentrated more near $z=0$ as $n$ increases (Figure \ref{fig:1D_rho}).

\subsection{Average gravitational amplification}

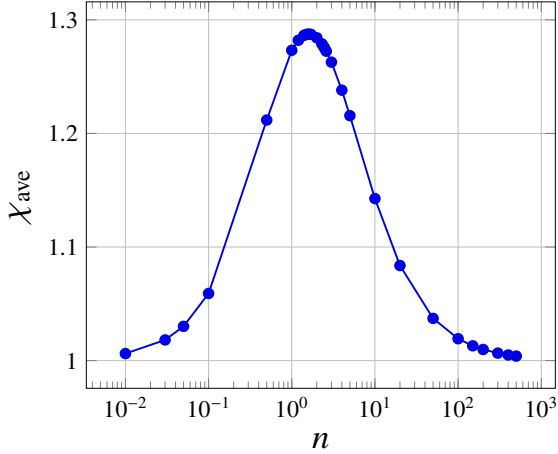
\begin{figure}
	\centering
	\begin{tikzpicture}[scale=0.9]
		\begin{semilogxaxis}[ grid=major,xlabel=\Large{$n$},ylabel=\Large{$\chi_\text{ave}$}, legend pos=north east]
			\addplot [color = bblue ,mark=*,thick,solid ] table[x index=0, y index=6]{Poly1D_n_rM_Mtot_EstM_rEstM_EstEff_EstAVE.txt}; 
		\end{semilogxaxis}
	\end{tikzpicture}
	\caption{Average gravitational amplification $\chi_\text{ave}$ against  polytropic index $n$ for the polytropic sheet models.}
	\label{fig:1D_Mtot_EstEff}
\end{figure}

The numerical value of $\chi_\text{ave}$ is shown in Figure \ref{fig:1D_Mtot_EstEff} against polytropic index $n$. As easily expected, the polytrope of $n=1$ takes a relatively high value of $\chi_\text{ave}$ since it does not depend on its unperturbed mass distribution and system size. The maximum of $\chi_\text{ave}$ occurs when $n\simeq 1.6$. This would be the outcome of the two effects; a moderately high index $n$ and slowly decaying density $\rho_\text{o}(z)$ in the perturbed density $(\propto n(-\phi_\text{o}(z))^{n-1}=n \rho_\text{o}^{1-1/n}(z))$ in equation \eqref{LE_1D_perturb_expan_delPhi}. For this, we first assume from the result of Section \ref{sec:n_0} that if the perturbed density  is close to zero then the amplification is not effective. For example, for very low $n(\ll 1)$, mass density distribution slowly diverges at large distances, but the perturbed density is still low at every height $z$ due to the low $n$. For higher $n(\gg 1)$, $\chi(z)$ can reach large values, but the average is low because of large polytrope height and rapid decay in the unperturbed density. We hence may expect the perturbed density to be more effective for amplification when $n$ is the order of 1.

\subsection{Shortening of the polytropic sheets}

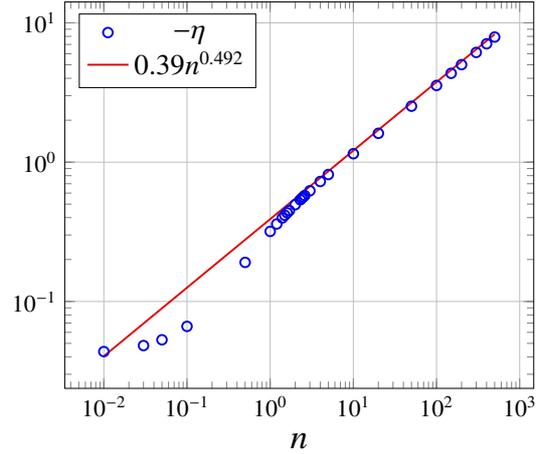
\begin{figure}
	\centering
	\begin{tikzpicture}[scale=0.9]
		\begin{loglogaxis}[ grid=major,xlabel=\Large{$n$}, legend pos=north west]
			\addplot [color = bblue ,mark=o,thick,solid, only marks] table[x index=0, y index=3]{Poly1D_n_Delx_ref1_Eta_ref2_epsil_ref3.txt}; 
			\addlegendentry{\large{$-\eta$}}
			\addplot [color = bred ,mark=no,thick ] table[x index=0, y index=4]{Poly1D_n_Delx_ref1_Eta_ref2_epsil_ref3.txt};
			\addlegendentry{\large{$0.39n^{0.492}$}}
		\end{loglogaxis}
	\end{tikzpicture}
	\caption{Negative of ratio $\eta$ of $\delta z$ to $z_\text{M}$ against  polytropic index $n$ for the polytropic sheet models. A guideline is depicted together as an approximation of $\eta$ for high index $n$.}
	\label{fig:1D_Eta}
\end{figure}

The \emph{bare} gravitational field due to test mass is constant $(=1/2)$ throughout polytropic sheets. The higher sheets can be deviated more largely from the original to the new equilibrium position because of low pressure. Polytropes with higher polytropic index hence experience more significant shortening in the $z$-direction (Figure \ref{fig:1D_Eta}). The ratio $\eta$ is well approximated by $\sim n^{0.492}$ for high indexes. This means that we need a very small value of $\epsilon$ for a high polytropic index. For example, to achieve 1$\%$ of shrinkage, or $\epsilon \delta z/ z_\text{M}=0.01$, the necessary value of $\epsilon$ is $\sim0.01 n^{-0.492}$; $\epsilon\sim10^{-5}$ for $n=500$. It is obvious that the linear approximation of potential $\phi(z)$ may break down for high $n$. This break-down could be reasonable by seeing the expanded form of $\phi(z)$ in equation \eqref{Eq.phi_expan}. The perturbation potential $\epsilon \delta \phi(z)$ increases like $\epsilon z/2$ as $z$ increases while $\phi_\text{o}(z)$ decreases. The linearization would successfully apply to polytropes with shorter heights. To hold the linear approximation for any $n$, polytropes must embedded in a pressurized medium. 

%% file: Section5_conclusion.tex
\section{Conclusion}\label{sec:conclusion}

The present work discussed gravitational polarization of fields induced by test mass sheet instantaneously placed at $z=0$ in polytropic sheets of $0\leq n<\infty$. It showed that gravitational amplification occurs even in a \emph{finite} extent of equilibrium self-gravitating systems in addition to previously reported infinite collisional and collisionless systems.  We first obtained analytical results for polytropic sheets of $n=0$ and $n=1$. The former model is a special case in which neither system shrinkage nor gravitational amplification does occur. The latter model is the simplest model that possesses basic features of gravitational amplification. It also helps confirm our numerical results.

Our numerical results showed that gravitational amplification $\eta$ in polytropic sheets are qualitatively the same as those in collisionless isochrone, isothermal sheet, and isothermal cylinder. The amplification is one at $z=0$, gets greater with height, and reaches its maximum value. After passing the maximum, it then decreases with height and equals one at the maximum height.  We found in the polytropic sheets that the maximum of $\eta$ increases with polytropics index $n$ and approaches that of the isothermal sheet model. The height at which the maximum occurs gets shorter for a higher polytropic index. We also found that the average of gravitational amplification is maximized around at $n=1.6$ and it gets weaker as $n$ gets greater or less.  Lastly, we computed the shortening of the polytropic sheets. The ratio of the height change to maximum height becomes greater with index $n$. It is approximately proportional to $n$ when $n\gg1$. This analysis provided a constraint on the present linear approximation.

In the future paper, we will extend the present work to nonlinear cases. We may also apply the present method to polytropic cylinder and sphere. Especially, the latter model is unstable to radial perturbation. We will confine the isothermal sphere in a rigid wall or by a pressurized medium or polytrope. The present formulation, especially for $n=0$, can be readily arranged for such confined systems. Another interesting extension work would be to move test mass from the original equilibrium polytrope to the center of gravity rather than adding extra test mass. Such a process was already discussed in \cite{Gilbert_1970} for collisionless isochrone. The discussion can be useful to understand the discreteness of self-gravitating systems; it will help understand how the ``statistical term \citep{Gilbert_1968}'' affects gravitational fields due to test mass and how the fields disappear near the surface of self-gravitating systems.